\documentstyle[aps,prl,twocolumn,epsf,rotate]{revtex}
\DeclareMathAlphabet{\foo}{U}{msb}{m}{n}
\begin{document}
\draft
\author{V.V. Sokolov$^{a}$, O.V. Zhirov$^{a}$,
        D. Alonso$^{b}$, and G. Casati$^{c\,\ast}$}
\address{$^a$ Budker Institute of Nuclear Physics,
        630090 Novosibirsk, Russia}
\address{$^b$ Departamento de F\'\i sica Fundamental y Experimental, La
Laguna 38203, 
Tenerife, Spain}
\address{$^c$ International Center for the Study of Dynamical Systems,
         22100 Como,  
and Istituto Nazionale di Fisica della Materia and INFN, Unita'di Milano,
Italy}
\title{Quantum Resonances of Kicked Rotor and $SU(q)$ group}
\date{\today}

\twocolumn[\hsize\textwidth\columnwidth\hsize\csname
@twocolumnfalse\endcsname

\maketitle

\begin{abstract}
The quantum kicked rotor (QKR) map is embedded into a continuous
unitary transformation generated by a time-independent quasi-Hamiltonian.
In some vicinity of a quantum resonance of order $q$,  we relate the 
problem  to the {\it regular} motion along a circle in a $(q^2-1)$-component 
inhomogeneous ``magnetic" field of a quantum particle with $q$ intrinsic degrees of 
freedom described by the $SU(q)$ group. This motion is in parallel with the 
classical phase oscillations near a non-linear resonance.
\end{abstract}
\vskip 1pc]

For years, Chirikov's standard quantum map \cite{CCFI79}, i.e. a planar
quantum rotor driven by very short periodic kicks,  served as a
cornerstone model for many investigations of onset and manifestations of
dynamical chaos in quantum systems. Indeed, the standard map provides a
local description for a large class of dynamical systems and it is of quite 
general interest to understand its properties. The unitary Floquet
transformation $U$ which evolves  the wave function $\psi(\theta)$ over 
each kick period is given by:
\begin{equation}\label{U}
\textstyle
U = U_r\cdot U_k\equiv\exp\left(-\frac{i}{2}T {\hat m}^2\right)\cdot
\exp\left(-ik\cos\theta\right)
\end{equation}
and consists of the operator $U_k$ of a kick with strength $k$ and a
consecutive free rotation $U_r$ during the time $T$. Here ${\hat m}=
-id/d\theta$ and we put $\hbar=1$. 

In the classical limit the motion depends on the single parameter $K=Tk$. 
When this parameter exceeds unity the classical dynamics yields unrestricted 
diffusive growth of the mean kinetic energy.  The quantum dynamics is 
appreciably richer \cite{Izr90}. In particular the phenomenon of dynamical 
localization has been discovered \cite{CCFI79}, namely the quantum motion 
mimics the classical diffusive behaviour only up to a time $t_R \sim k^2$ 
after which it enters a stationary oscillatory regime. 

However, at present, we are still very far from a satisfactory understanding
of this simple model of quantum chaos. The main obstacle for the analytical 
and  numerical investigations are the so called {\it quantum resonances} 
which take place at a fixed value of $k$ for the everywhere dense set of the 
rational values $\varsigma=T/4\pi =p/q$, the integer numbers $p$ and $q$ 
being mutually prime \cite{IzSh79,Izr90}. In such a regime, the mean kinetic 
energy increases quadratically for asymptotic large times. Quantum resonances 
may strongly influence the motion in some vicinities around them. In other 
words they have certain widths. It is in this feature the main difference between 
the QKR and disordered quantum systems and the main reason for the failure of 
analytical tools in particular those which use supersymmetric technique 
\cite{AltZir96,CIS98}. Therefore any attempt to improve our understanding 
of quantum chaotic motion must confront with the so far unsolved core problem 
of the motion in vicinities of resonances.

In this paper we develop a general approach to the problem of the motion
in a vicinity of a quantum resonance with arbitrary order $q$. We show that 
the most important role is played by the resonances with small $q$: for 
finite regions around them the quantum motion is explicitly shown to be regular
and dominates the motion for all $\varsigma$ values inside these regions.
More precisely, the motion is described, for large though finite times, by 
a conservative Hamiltonian with one rotational degree of freedom and with 
a discrete spectrum.  On the contrary, when the resonance order is large enough
the resonant quadratic growth appears only in the remote time asymptotic 
and during a long time the motion reveals universal features characteristic 
of the localized quantum chaos. We demonstrate on a specific example 
that our approach allows to predict the quantum evolution with great 
accuracy for long time (see Fig.1) even though the corresponding classical 
motion is chaotic and exponentially unstable.

Basically, we use the same idea of embedding the map (\ref{U}) into a
continuous unitary transformation which has been used before \cite{Sok67} 
in the study of the classical counterpart of (\ref{U}).  The Floquet 
operator is represented in the form $U=\exp(-i{\cal H})$, so that the motion 
appears as a conservative evolution in a {\it continuous} time $t$ of
a system with one rotational degree of freedom described by the effective 
time-independent {\it quasi-Hamiltonian} ${\cal H}$. The latter cannot 
generally be found in a closed form. Rather, it is formally expressed as an 
infinite sum of successive commutators. However, when the condition of 
a quantum resonance is fulfilled it simplifies enormously \cite{IzSh79} and 
reduces to a pure phase (or local gauge) transformation from the unitary 
unimodular group $SU(q)$. Moreover, in some vicinities of quantum  
resonances, the sum of commutators can be truncated thus allowing to obtain 
new insight on the nature of the motion.

Before turning to the consideration of the general case, we illustrate our
approach on the simplest and strongest resonances $q=1, 2 ,4$ (see also 
\cite{IzSh79}). For the main resonance $q=1$ the rotation operator is 
equivalent to identity. The Floquet transformation (\ref{U}) is then simply 
$e^{-iv(\theta)}$, where $v(\theta)=k\cos\theta$ is the interaction
potential. After $t$ repeatitions of this transformation, an initially 
isotropic wave function gets $\sim k t$ harmonics. This number is naturally 
measured by the operator ${\hat m}(t)\equiv e^{ivt}{\hat m} e^{-ivt}={\hat m}-
v'(\theta)t$. This yields the increase law $E_k(t)\equiv \langle\Delta {\hat
m}^2(t)\rangle/2=r\,t^2$ with the {\it resonant growth rate} $r(k)=\langle
v'{\,^2}\rangle=k^2/4$ for the kinetic energy. The symbol $\langle...\rangle$ 
stands here for the average over the angle $\theta$. 

For the other two resonances $q=2,4$ the rotation operator $U_r$ has only 
two eigenvalues $(1, -1)$ and $(1, -i)$ respectively. The corresponding 
eigenfunctions $\psi^{(\pm)}(\theta)$ are periodic and contain only even 
(odd) harmonics for the first (second) eigenvalue in each case. By writing the
wave function in the form of a spinor $\Psi(\theta)$
\begin{equation}\label{Psi2}
\Psi^{T(\equiv transposed)}(\theta) = 
\left(\psi^{(+)}(\theta)\,\,\,\,\psi^{(-)}(\theta)\right)
\end{equation}       
the rotation is described by a matrix operator ${\foo{U}}_r=e^{-i\alpha_r(\sigma_3-1)}$
with $\alpha_r=\pi/2$ for $q=2$ and $\alpha_r=-\pi/4$ if $q=4$. The kick 
matrix  has in both cases the same form $e^{-iv\sigma_1}$. Here $\sigma_j$ are
$2\times 2$ Pauli matrices. Simple manipulations cast, up to an irrelevant
constant phase, the Floquet operator, that acts on the spinor $\Psi(\theta)$, 
into the form of a transformation from the $SU(2)$ group
\begin{equation}\label{U2,4}
{\foo U}^{(res)}_q = \exp\left(-iw\,{\bf n}\cdot\mbox{\boldmath 
$\sigma$}\right)
\equiv \exp\left(-i\tilde{\cal H}^{(res)}(\theta)\right),
\end{equation}
where for $q=2$, $w=\pi/2$, and ${\bf n}=(0, \sin v, \cos v)$ while for
$q=4$, $w=\arccos(\cos v/\sqrt{2})$ and ${\bf n}= (1+\sin^2v)^{-1/2}
(\sin v,-\sin v, \cos v)$. 
The hermitian operator in the exponent is equivalent to the Hamiltonian of 
the spin 1/2 in a magnetic field oriented along the vector {\bf n}.

The unified representation (\ref{U2,4}) allows us to calculate in closed
form the evolution of the angular momentum operator 
${\foo M}=diag({\hat m},{\hat m})$  
$$\Delta {\foo M}(t) = e^{iw\,{\bf n}\cdot\mbox{\boldmath $\sigma$}\,t}
{\foo M} e^{-iw\,{\bf n}\cdot \mbox{\boldmath $\sigma$}\,t}-{\foo M}$$ 
\begin{eqnarray}\label{M(t)2}
&=& -\int_0^t d\tau \left[w'{\bf n}\cdot\mbox{\boldmath $\sigma$} +
w\,{\bf n'}\cdot e^{-iw\,{\bf n}\cdot\mbox{\boldmath {\tiny
$\Sigma$}}\,\tau}
\mbox{\boldmath $\sigma$}\right]     \\ \nonumber
&=& -\left(w't\,{\bf n}+\frac{\sin 2wt}{2}\,{\bf n'}+\frac{1-\cos 2wt}{2}
{\bf n'}\times{\bf n}
\right)\cdot\mbox{\boldmath $\sigma$}.  \\ \nonumber
\end{eqnarray}
The 3 matrices $\Sigma_a,\, a=1,2,3$ are  generators of the adjoint
representation of the $SU(2)$ group while the vectors ${\bf n}, {\bf n'}$ 
and ${\bf n'}\times{\bf n}$ define an orthogonal basis in this space. 
At a given angle, the evolution (\ref{M(t)2}) consists of two different 
contributions. The linearly growing part appears because of inhomogeneity 
of the magnetic field while the periodic part is due to the spin rotation 
in this field. The corresponding kinetic energy evolution reads
\begin{equation}\label{varE2}
E_k(t) = r\,t^2+\chi(t)
\end{equation}
where 
\begin{equation}\label{2,4}
r(q;k)=\frac{1}{2}\langle (w')^2\rangle, \qquad
\chi(q;k;t)=\frac{1}{2}\langle{\bf n'}^{ \, 2}
\sin^2w t\rangle .
\end{equation}

When $q=2$, the quantity $w=\pi/2$ is independent of the angle $\theta$ 
so that the quadratically growing term disappears and the function 
$\chi(t)$ is purely periodic
\begin{equation}\label{q=2}
\textstyle
E_k(t) =
\frac{1}{4}k^2\sin^2(\frac{1}{2}\pi t).
\end{equation}
Only the spin rotation contributes and the energy  jumps between
values 0 and $k^2/4$ when the time $t$ runs over integers. On the other 
hand, for $q=4$ both contributions exist. The function $\chi(t)$ fluctuates 
with time approaching as $const/\sqrt{t}$ the finite positive value
\begin{equation}\label{aschi}
\chi_{\infty} = \frac{k^2}{2\pi}\int_0^{\pi}\!\! d\theta
\sin^2\theta \left[1+\sin^2v\right]^{-2}.
\end{equation}

Let us now  detune slightly from the exact resonance, $T=T^{(res)}
+\kappa$. The Floquet operator looks then as 
${\foo U}_{p,q}(\kappa)=
\exp\left(-\frac{i}{2}\kappa {\foo M}^2\right){\foo U}_{p,q}^{(res)}=
\exp\left(-\frac{i}{\kappa} {\cal H(\kappa)}\right)$. Representing the
quasi-Hamiltonian as
\begin{equation}\label{H}
{\cal H} = \kappa\tilde{\cal H}^{(res)}(\theta) +\kappa^2 Q(\kappa),
\end{equation}
we come to the condition
\begin{eqnarray}\label{Q}
&&\textstyle
\exp\left(-\frac{i}{2}\kappa {\foo M}^2\right)= \nonumber\\
&&\quad T^*\exp\left\{-i\kappa\int_0^{1}d\tau\,
e^{-i\tilde{\cal H}^{(res)}\tau}\,Q(\kappa)\,
e^{i\tilde{\cal H}^{(res)}\tau}\right\}.
\end{eqnarray}
The symbol $T^*$ indicates the antichronological ordering. One can formally 
solve this equation by expanding the operator $Q(\kappa)$  over the 
small detuning $\kappa$. The effective quasi-Hamiltonian resulting from 
such a procedure appears in the form of the series
\begin{equation}
\label{Hser}
{\cal H} = \kappa \tilde{\cal
H}^{(res)}(\theta)+\frac{1}{2}\left\{J, F_1
(\theta)\right\}_++\frac{1}{2}\, JF_2(\theta)J+...
\end{equation}
which is, in particular, an expansion in powers of the angular momentum 
$J=\kappa {\foo M}$ in which the zero-order term generates the resonant
Floquet transformation. All operators in (\ref{Hser}) are one or two 
dimensional matrices depending on the order $q=1$ or 2, 4. Being a factor 
in front of the angle derivative, the detuning $\kappa$ plays in (\ref{Hser}) 
the role of the dimensionless Planck's constant. Keeping only the terms
written in eq. (\ref{Hser}) explicitly, the quasi-Hamiltonian ${\cal H}$, when
$q=2, 4$, is formally equivalent to the Hamiltonian of a ``particle'' with the
spin 1/2 which moves along a circle in an inhomogeneous magnetic field. 
The term linear in the angular momentum $J$ mimics a sort of spin-orbital 
interaction. Due to the periodic boundary condition, the eigenvalue spectrum 
$\{\epsilon\}$ of ${\cal H}$ is discrete so that the detuning restricts the
growth of the energy to a finite maximal value. This justifies the expansion 
over the angular momentum. 

Comparing the results predicted by the quasi-Hamiltonian  with
exact numerical simulations of the original quantum map (\ref{U}),  we 
found that already the approximation (\ref{Hser}) describes quite satisfactorily 
not only the cut-off of the initial resonant growth (when it exists in the 
resonance regime) and the mean value of the kinetic energy after the growth
has been saturated but also delicate features of  very irregular quantum 
fluctuations in the plateau region.  The term quadratic in the angular momentum $J$ 
turns out to be of the principal importance. 

As a typical example we show in Fig.1 the results for the resonance $q=4$.  
The points are numerical simulations with the exact Floquet operator (\ref{U})
(only each 4th point is kept in the main part and 500th in the inset) while 
the solid line corresponds to the evolution described by the time-independent 
quasi-Hamiltonian (\ref{Hser}).   The condition under which 
the influence of the lowest omitted correction is weak gives an estimate of the width 
$\Delta\kappa$ of a resonance \cite{SZAC}. This leads to $\Delta\kappa\sim 1/k$ for the
resonances $q=1,2$. The resonance $q=4$ is much narrower and $\Delta\kappa 
\propto 1/k^2$. Both estimates are in agreement with the numerical data.
Outside these intervals the expansion becomes transparently divergent. 
In Fig.2 the transient region of the motion near the resonance $q=4$ as it
comes from the exact numerical simulations is presented. Two quite 
different regimes are clearly seen. The crossover from the regular (lower plateau 
which is  well described in terms of the quasi-Hamiltonian) to chaotic (upper 
line) regimes takes place in a rather narrow interval of the detuning.  To 
explore the regularity domain and the adjacent region, we have fitted the height 
$E_{pl}$ of the plateau in Fig.2 as a function of the detuning $\kappa$, 
(see Fig.3). The plateau height scales  as $\kappa^{-1}$ in the regularity
domain $(\kappa\leq 10^{-4})$, while in the ``quantum chaos" region 
$(\kappa\geq10^{-3})$, the plateau height is scattered around the expected 
localization value $\propto l^2$.  Note that the resonance essentially
suppresses the diffusion in the intermediate region as well. Higher 
corrections to eq. (\ref{Hser}) may be relevant in this domain.

Rigorously speaking, an expansion of  the kind (\ref{Hser}) cannot, in 
spite of the satisfactory agreement with the exact numerical solution, 
converge. Indeed, an infinite number of quantum resonances of large 
order hit the domain $\Delta\kappa$ of influence of a strong resonance 
with a small order. In such a resonance of high order q, the Floquet 
operator is a  $SU(q)$ transformation
\begin{equation}\label{SU}
{\foo U}_{p,q}^{(res)} =
\exp\left(-iw\,{\bf n}\cdot \mbox{\boldmath $\lambda$}\right);
\quad {\bf n}^2 = 1
\end{equation}
which is a plain extension of  eq. (\ref{U2,4}). Here the matrices 
$\lambda_a$ are the generators of the $q$-dimensional fundamental 
representation of the group, ${\bf n}\cdot \mbox{\boldmath $\lambda$}
\equiv\sum_a n_a \lambda_a$ and ${\bf n}$ is a unit vector in the 
$(q^2-1)$~-dimensional adjoint space. The transformation (\ref{SU})
depends on the angle $\theta$ via the periodic functions $w(\theta)$ and
${\bf n}(\theta)$ which in turn are expressed in terms of the kick 
potential $v(\theta)$. These functions satisfy a system of $q^2-1$ transcendental 
equations which, generally, cannot be solved analytically. Nevertheless, 
some generic information can be obtained even without knowing the explicit solution. 
Indeed, the evolution of the $q\times q$ angular momentum matrix is 
described, similarly to eq. (\ref{M(t)2}), by 
\begin{equation}\label{M(t)q}
\Delta {\foo M}(t) = \left[{\bf M}^{(0)}\,t+ {\bf
M}^{(1)}(t)\right]\cdot\mbox{\boldmath$\lambda$}.
\end{equation}
The matrices ${\bf M^{(0)}}$ are expressed in terms of the q-1 zero modes 
of the matrix $({\bf n}\cdot\mbox{\boldmath $\Lambda$})$ which acts in the
$(q^2-1)$-dimensional adjoint space,  while the quasiperiodic (as long as 
the angle $\theta$ is kept fixed) matrix ${\bf M^{(1)}}(t)$ is formed by the
$q(q-1)/2$ pairs of mutually conjugate modes with finite eigenvalues
$\pm\xi_a$ \cite{SZAC}. 
These two matrices determine respectively the resonant growth rate
\begin{equation}\label{rq}
\textstyle
r(p,q;k) = \frac{1}{2}\sum_{a}\eta_a\langle[M_a^{(0)}]^2\rangle
\end{equation}
and the asymptotic value
\begin{equation}\label{sat}
\textstyle
\chi_{\infty} = 2\sum_{\{\xi_b>0\}}\langle\xi_b^{-2}
|{\bf n'}\cdot \mbox{\boldmath $\chi$}^{(b)}|^2\sum_a\eta_a
|\chi_a^{(b)}|^2\rangle.
\end{equation}
The quantities $\eta_a$ in this formulae select the (2q-1)-dimensional
{\it active} subspace of the adjoint space, which is reachable for the 
motion under the isotropic initial conditions. A small detuning from 
the considered resonance kills the unrestricted growth created by the 
resonance itself and gives rise to the $q\times q$  quasi-Hamiltonian 
matrix of the same structure as in (\ref{Hser}).  The motion looks like that of a
``particle" with q intrinsic degrees of freedoms described by the $SU(q)$
group, along a circle in a $(q^2-1)$-component ``magnetic" field.

Returning to the convergence problem, one sees that, independently of the 
number of the corrections taken into account, the quasi-Hamiltonian
approach fails to reproduce the unrestricted resonant growth in the 
resonant points within the domain of the influence of a strong resonance (i.e. with 
a small order) around which the expansion is performed. Indeed, the spectrum 
of the quasi-Hamiltonian is always discrete whereas the resonant growth 
implies continuous spectrum. 

However, as confirmed by numerical data, the growth rate $r(p,q;k)$ 
is exponentially small when $q$ noticeably exceeds the localization 
length \cite{Chir}. Therefore, the resonant growth reveals itself only on 
a very remote time asymptotics. Qualitative arguments presented in 
\cite{Chir89} connect this fact with the exponentially weak overlap of 
the neighboring localized parts of the globally delocalized quasienergy 
eigenfunctions. Exponential effects of such a kind, which are characteristic 
of the tunneling, are well known to be beyond the reach of  perturbation 
expansions. That is why they cannot be described in the framework of the  
quasi-Hamiltonian method. The latter reproduces well only those features 
of the motion, which are determined by the discrete component of the
quasienergy spectrum, in particular, the function $\chi(t)$ which attains 
its asymptotical value (\ref{sat}) much faster. Therefore inside the width 
of a strong resonance the time dependence of these functions is dictated 
for all weak ones by their strongest brother.  

On the other hand, if the order is very large, $q\gg l$, and the resonance
lies in the region of typical irrationals being far from all strong
resonances, already a very small detuning suffices for killing the quadratic 
growth with exponentially small rate \cite{ftnt}. At the same time,  such a 
shift does not influence the function $\chi(t)$ which reproduces on a large 
(though finite) time scale characteristic features of the ``localized quantum 
chaos'' \cite{Izr90}.  

\begin{figure}[h]
\unitlength 1cm
\begin{picture}(8, 3.7)
\epsfxsize 8cm
\epsfysize 4cm
\put(0,0.){\epsfbox{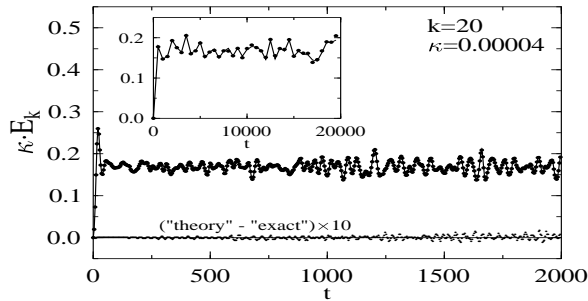}}
\end{picture}
\unitlength 1bp
\caption{
The energy $E$ times $\kappa$ versus the number of  kicks for 
$q$=4. Deviation of the theory from the exact map is indicated at 
the bottom.
}
\label{fig1}
\end{figure}

\begin{figure}[t]
\unitlength 1cm
\begin{picture}(8,3)
\epsfxsize 7.5cm
\epsfysize 3.5cm
\put(0,0.1){\epsfbox{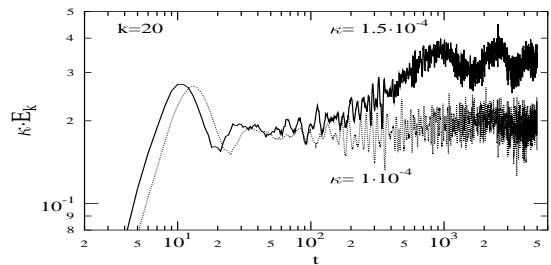}}
\end{picture}
\unitlength 1bp
\caption{
The crossover region near the resonance $q$=4. Double log. scale
is chosen. The resonant behavior on the initial stage is seen in both 
cases.
}
\label{fig2}
\end{figure}

\begin{figure}[t]
\unitlength 1.1cm
\begin{picture}(8,3)
\epsfxsize 7.5cm
\epsfysize 3.8cm
\put(0,0.1){\epsfbox{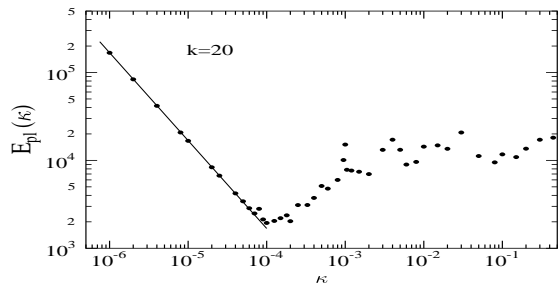}}
\end{picture}
\unitlength 1bp
\caption{
The plateau height $  E_{pl}$ versus detuning $\kappa$ near
the resonance $q$=4. The solid line corresponds to the theoretical
relation $E_{pl}\propto 1/\kappa$.
}
\label{fig3}
\end{figure}

The behavior becomes more complicated if a number of resonances with
comparable and moderate orders are neighbouring and their domains overlap.
The expansion near one of them forms a plateau which lasts until the 
quadratic growth in a next resonance of the same strength reveals itself so
that the original expansion fails. However, the expansion near the new resonance 
cuts off this growth and forms a higher plateau until a next resonance comes to 
the action. Such a pattern of repeatedly reappearing regimes of resonant growth
has been discovered in \cite{CFGV86}

In conclusion, the concept of the time-independent quasi-Hamiltonian of a
quantum map is proposed. The motion of the quantum kicked rotor in
a quantum resonance of order $q$ is exactly described by a continuous
transformation from the $SU(q)$ group. The motion in some vicinity of this
resonance is proved to be similar to that of a quantum particle with $q$
intrinsic degrees of freedom along a circle in an inhomogeneous 
$(q^2-1)$-component ``magnetic" field. The latter is in analogy with the 
classical phase oscillations near a non-linear resonance. The resonances 
with sufficiently small $q$ master the motion in finite domains near the 
resonance points though the widths of this domains rapidly diminish 
with the order.

We are very grateful to B.V. Chirikov for making his results available to
us prior to publication, F.M. Izrailev for many useful remarks and D.V. 
Savin for advices. V.S. thanks the ICSDS for a kind hospitality during his stay 
in Como supported by the Cariplo Foundation. The financial support from RFFR,
grant No 99-02-16726 (V.S. and O.Z.) and the EU program Training and 
Mobility of Researches, contract No ERBFMRXCT960010 and Gobierno 
Aut\'onomo de Canarias (D.A.) are acknowledged with thanks.

\mbox{\ }


\vspace{-1cm}

\begin{thebibliography}{99}
\vspace{-1.5cm}
\bibitem[\ast]{e-mail} e-mail: casati@fis.unico.it
\bibitem{CCFI79} G. Casati, B.V. Chirikov, J. Ford, and F.M. Izrailev,
Lecture Notes in Physics {\bf 93} 334 (1979).
\bibitem{Izr90} F.M. Izrailev, Phys. Rep. {\bf 196}, 299 (1990).
\bibitem{IzSh79} F.M. Izrailev and D.L. Shepelyansky,
Sov. Phys. Dokl. {\bf 24}, 996 (1979); Sov. Theor. Math. Phys {\bf 43},
553 (1980).
\bibitem{AltZir96} A. Altland and M.R. Zirnbauer,
Phys. Rev. Lett. {\bf 77}, 4536 (1996).
\bibitem{CIS98} G. Casati, F.M. Izrailev, V.V. Sokolov.
Phys. Rev. Lett. {\bf 80}, 640 (1998).
\bibitem{Sok67} V.V. Sokolov,
Sov. Journ. Theor. Math. Phys. {\bf 67}, 223 (1986); F.M. Izrailev and
V.V. Sokolov, Phys. Lett. {\bf A112}, 254 (1985).
\bibitem{SZAC} V.V. Sokolov et al: in preparation.
\bibitem{Chir} B.V. Chirikov et al: in preparation.
\bibitem{Chir89} B.V. Chirikov, in {\it Chaos and Quantum Physics},
Proc. of the Les Houches Summer School, Session LII, edts.
M.J. Giannoni {\it et al.}, (Elsevier, Amsterdam, 1991), p.443.
\bibitem{ftnt} The resonant growth may be also removed by  
putting the system on a torus in the momentum space \cite{Izr90}.
\bibitem{CFGV86}  G. Casati, J. Ford, F.Vivaldi and I. Guarneri,  Phys. 
Rev. {\bf A 34}, 1413 (1986).
\end{thebibliography}
\end{document}